\documentclass[showkeys]{revtex4}
\usepackage[cp866]{inputenc}   
\usepackage[T2A]{fontenc}      
\usepackage{amsmath}           
\usepackage{amssymb}           
\usepackage[dvips]{graphicx}   
\usepackage{euscript}

\usepackage{amssymb}
\usepackage{amsmath}
\usepackage{euscript}
\usepackage[dvips]{graphicx}
\usepackage{epsfig}
\begin{document}

\title{Geodesics on deformed spheres asymptotically described using the Funk transform}

\author{D.O.\,Sinitsyn}

\affiliation{Department of Mechanics and Mathematics\\
M.V. Lomonosov Moscow State University\\
Moscow, Russia}

\begin{abstract}

We consider geodesics on the surfaces obtained by weak deformations of the standard 2D-sphere.
The dynamics of a particle on the surface can be asymptotically described by the averaged evolution
of the particle's angular momentum. It is shown that the system describing this evolution
has a Hamiltonian, which is obtained by applying the Funk transform to the function
defining the deviation of the surface from the standard sphere.
This system has the 2D-sphere as its phase space, so it is integrable
and its trajectories admit of topological description in terms of its phase portrait on the sphere.

\end{abstract}


\date{\today}

\maketitle

\section{Introduction}

The problem of finding geodesics on a surface is a classical subject of differential geometry and analytical mechanics.
Generally the exact analytical treatment of the dynamics equations gets quite involved even for
comparatively simple surfaces such as ellipsoid (the well-known result of Jacoby, \cite{Jac}).
On the other hand, if we are interested in the qualitative topological structure of the system's
trajectories and their large-scale properties, exact solutions are not necessary.
For such purposes we can apply asymptotic methods, which provide a general picture of the system's dynamics.

In the present work we consider surfaces that are obtained by a small deformation of the standard 2D-sphere.
We perform an asymptotic Hamiltonian reduction of the dynamics of a particle on such a surface
to a Hamiltonian system with one degree of freedom. The latter system is always integrable
and admits of visual description by means of phase portraits.
The idea underlying the reduction was introduced in \cite{GS}, where a special class of
deformed spheres was analyzed in detail.
The key observation is that for small periods of time the trajectory of the particle is close
to the exact solution of the unperturbed problem, i.e. a geodesic on the standard sphere -- its great circle.
So, the dynamics of the particle can be approximately described by how the position of
this great circle slowly changes in time.

Here we show that the result holds for any small
smooth deformation of the sphere and that the procedure of obtaining the Hamiltonian
of the reduced system consists in applying the Funk-Minkowski-Radon transformation (see \cite{Gelf}) of functions on the 2D-sphere
to the function defining the difference of the surface from the sphere.

\section{Hamiltonian reduction}

\subsection{Dynamics equations}

We consider the motion of a particle on a surface defined by the equation
$$
\varphi(\vec x) = 0
$$
where
$$
  \varphi(\vec x) = \sum_{i=1}^{3} x_i^2 - 1 + \varepsilon \,\psi(\vec x), \quad \varepsilon \ll 0.
$$
So, the surface is obtained by a small perturbation $ \psi(\vec x) $ of the standard sphere.

The motion of a particle with unit mass on the surface can be described by the Lagrange equations
of the first kind:
  \begin{equation}
    \ddot{\vec x}
    = \lambda \,
    \frac{\partial \varphi}
     {\partial \vec x}   \mbox{.}
    \label{2Newton}
  \end{equation}
The Lagrange multiplier $ \lambda $ can be found explicitly
  \begin{equation}
   \lambda = - \frac{\dot{\vec x} \cdot
          \displaystyle
          \frac{\partial^2 \varphi}{\partial
                \vec x^2} \cdot
           \, \dot{\vec x}
          }
         {\left( \displaystyle
         \frac{\partial \varphi}{\partial \vec x}
          \right)^2}
= - \frac{\dot{\vec x}^2
+ \varepsilon \, \dot{\vec x} \cdot \displaystyle \frac{\partial^2 \psi}{\partial \vec x^2} \cdot \, \dot{\vec x}}
{\left( \vec x + \varepsilon \displaystyle \frac{\partial \psi}{\partial \vec x} \right)^2}
\label{lamb}
\end{equation}
Here $\displaystyle \frac{\partial^2 \psi}{\partial \vec x^2}$ is the matrix of second derivatives of $\psi(\vec x)$:
$$
\left(\frac{\partial^2 \psi}{\partial \vec x^2}\right)_{ij} = \frac{\partial^2 \psi}{\partial x_i \partial x_j}.
$$

\subsection{Asymptotic descirption in terms of the momentum}

\begin{figure}
  \begin{center}
    \includegraphics[width = 150bp]{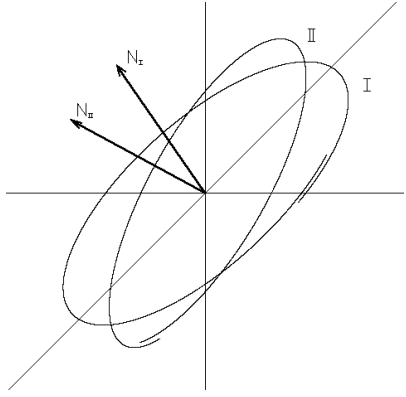}
    \caption{Two coils I, II of a trajectory on the surface $ \varphi(\vec x) = 0 $;
     vectors $ N_I $ and $ N_{II} $ are the normals to the planes of great circles approximating the coils}
    \label{fig1}
  \end{center}
\end{figure}

The trajectories of the particle have the following shape: small segments turn around the surface
close to a plane section of it -- a great circle (fig. \ref{fig1}). After a period of time this
great circle changes its position on the surface. So, we can approximately describe the motion
of the particle by tracing the slow change of the position of this great circle, thus
determining at every moment of time the approximate location of the current loop of the geodesic.

The position of a great circle can be described by a normal vector to it.
It is easy to see that the angular momentum of the particle
$$
\vec L = \vec x \times \dot{\vec x}
$$
provides such a normal vector. Its time derivative reads
\begin{equation}
\dot{\vec L} = \vec x \times \ddot{\vec x}
= \vec x \times \lambda \, \frac{\partial \varphi} {\partial \vec x}
= \lambda \, \vec x \times (\vec x + \varepsilon \frac{\partial \psi} {\partial \vec x})
= \lambda \,\varepsilon \, \vec x \times \frac{\partial \psi} {\partial \vec x}.
\label{mom}
\end{equation}
Taking into account here only the first order terms in $\varepsilon$, we can substitute $\lambda$ with
its zero order component obtained from~(\ref{lamb}): $\lambda_0 = -\dot{\vec x}^2 = const$ (the constancy follows
from conservation of the kinetic energy $E=\dot{\vec x}^2/2$). In what follows we assume the units to be chosen so that
$\dot{\vec x}^2 = 1$ and hence $\lambda_0 = -1$.

\subsection{Averaging the momentum equations}

We can take advantage of the fact that the system is close to an integrable one -- geodesics on the standard sphere.
This leads to the system having a fast variable -- the phase of the motion
along the current loop (close to a great circle), and slow variables defining the position of the loop.
The slow variables (components of the momentum) are first integrals for the unperturbed system.
So, we are in a position to apply the method of averaging, \cite{ArnMech}.
To derive an asymptotic description of the motion of the particle
in terms of the evolution of its angular momentum~$\vec L$, we average the above equation
over a period of the basic solution (for the unperturbed sphere) with a given (constant) momentum $\vec L$:
$$
    \vec x_{\vec L}(t) = \cos{t} \ \vec e_1(\vec L) + \sin{t} \  \vec e_2(\vec L),
$$
where vectors $\vec e_1(\vec L), \vec e_2(\vec L)$ form an orthonormal basis together with $ \vec e_3(\vec L) = \vec L / L $
(the exact choice of $\vec e_1(\vec L), \vec e_2(\vec L)$ does not affect the result of averaging).
This is the motion along the great circle perpendicular to the momentum $\vec L$.

\begin{figure}
  \begin{center}
    \includegraphics[width = 250bp]{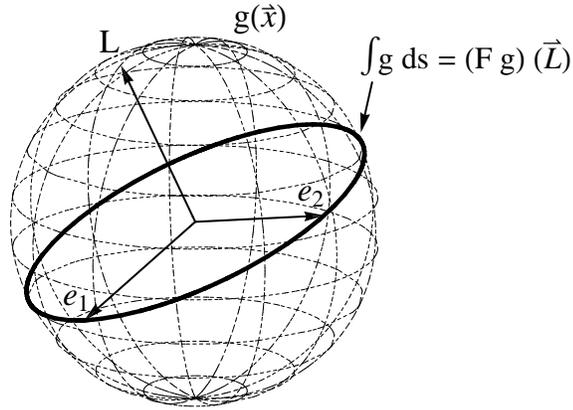}
    \caption{The Funk transform of a function $g(\vec x)$ on the unit sphere, taken at the point $\vec L$ of the sphere,
             is the integral of $g$ over a great circle perpendicular to $\vec L$.}
    \label{fig2}
  \end{center}
\end{figure}

So, we can see that the above averaging is equivalent to the {\it integration
of a function on the sphere over great circles}, i.e. the {\it Funk transform} (see \cite{Gelf}):
$$
(F g(\vec x)) (\vec L) = \int_0^{2 \pi} g\left(\cos{t} \ \vec e_1(\vec L) + \sin{t} \ \vec e_2(\vec L)\right) \, dt,
$$
where $\vec e_1(\vec L)$ and $\vec e_2(\vec L)$ are the above-mentioned unit vectors orthogonal to $\vec L$ and to each other
(fig. \ref{fig2}).

The Funk transform has the following basic properties:
\begin{enumerate}
\item Linearity:	$F(\lambda \, g + \mu \, h) = \lambda \, F(g) + \mu \, F(h).$ \newline
This follows from the linearity of integration.
\item Evenness of the image functions: $(F \, g)(-\vec L) = (F \, g)(\vec L)$ (so, the image of the transform is a function on
the projective plane). \newline
 This is due to the fact that $\vec L$ and $-\vec L$ are perpendicular to the same great circle.
\item Zero image of odd functions: if $g(-\vec x) = -g(\vec x)$ then $F \, g = 0$. \newline
Indeed, for such $g(\vec x)$ we have
$$
\begin{array}{lcl}
(F \, g(\vec x))(\vec L)
= \int_0^{\pi} g\left(\cos{t} \ \vec e_1(\vec L) + \sin{t} \ \vec e_2(\vec L)\right) \, dt
+ \int_{\pi}^{2 \pi} g\left(\cos{t} \ \vec e_1(\vec L) + \sin{t} \ \vec e_2(\vec L)\right) \, dt = \\
= \int_0^{\pi} g\left(\cos{t} \ \vec e_1(\vec L) + \sin{t} \ \vec e_2(\vec L)\right) \, dt
+ \int_{0}^{\pi} g\left(\cos(u+\pi) \ \vec e_1(\vec L) + \sin(u+\pi) \ \vec e_2(\vec L)\right) \, du = \\
= \int_0^{\pi} g\left(\cos{t} \ \vec e_1(\vec L) + \sin{t} \ \vec e_2(\vec L)\right) \, dt
+ \int_{0}^{\pi} g\left(-\cos{u} \ \vec e_1(\vec L) - \sin{u} \ \vec e_2(\vec L)\right) \, du = 0.
\end{array}
$$
\end{enumerate}

So, averaging of the equation for the momentum (\ref{mom}) gives
\begin{equation}
\begin{array}{lcl}
\displaystyle
\left\langle \dot{\vec L} \right\rangle
= - \varepsilon \left\langle \vec x \times \frac{\partial \psi} {\partial \vec x} \right\rangle_{\vec x \bot \vec L}
= - \frac{1}{2 \pi} \,\varepsilon
\int_0^{2 \pi} \vec x_{\vec L}(t) \times \frac{\partial \psi} {\partial \vec x}(\vec x_{\vec L}(t)) \, dt = \\ \\
\displaystyle
= - \frac{1}{2 \pi} \,\varepsilon
\left( F \left(\vec x \times \frac{\partial \psi} {\partial \vec x}\right) \right) \, (\vec L)
= - \frac{1}{2 \pi} \,\varepsilon
\left( F \circ \vec l \ \psi \right) \, (\vec L),
\label{momAv}
\end{array}
\end{equation}
where we have denoted the operator $\displaystyle \vec l = \vec x \times \frac{\partial} {\partial \vec x}$.

\subsection{The key commutation relation for the Funk transform}

Now we are going to derive the following general identity:
\begin{equation}
F \circ \vec l = \vec l \circ F,
\label{comm}
\end{equation}
where $ \vec l $ is the operator
$$
\vec l \, \psi(\vec x) = \vec x \times \frac{\partial \psi} {\partial \vec x}.
$$
For that end we employ the following property of the Funk transform.

\subsection{Commutation of the Funk transform with rotations}

The Funk transform commutes with the action of three-dimensional rotations.
Indeed, for any rotation $R \in SO(3)$ acting on functions as
$$
R\, \psi(\vec x) = \psi(R^{-1} \vec x)
$$
we have
\begin{equation}
(R \, (F \, \psi)) (\vec L)
= (F \, \psi) (R^{-1} \,\vec L)
= \int_{\vec x \bot (R^{-1}\,\vec L)} \psi(\vec x) \, ds
= \int_{(R\, \vec x) \bot \vec L} \psi(R^{-1} \, R \, \vec x) \, ds
= \int_{\vec y \bot \vec L} \psi(R^{-1} \vec y) \, ds
= (F \, R \, \psi)(\vec L),
\label{rot}
\end{equation}
i.e., we have obtained that
\begin{equation}
R \circ F = F \circ R.
\label{rot}
\end{equation}

\subsection{Proof of the key commutation relation}

Now we note that the components of the operator $\vec l$ are
the infinitesimal rotations:
$$
l_i = -\left(\frac{d}{d \alpha} R_i(\alpha)\right)|_{\alpha=0},
$$
where $R_i(\alpha)$ is the rotation about the axis with number $i$ through the angle $\alpha$.
E.g., for $i=3$ we have the rotation matrix
$$
R_3(\alpha)=
\begin{pmatrix}
\cos{\alpha} & -\sin{\alpha} & 0 \\
\sin{\alpha} & \cos{\alpha} & 0 \\
0            & 0            & 1
\end{pmatrix}
$$
and hence
$$
\begin{array}{lcl}
\displaystyle
\frac{d}{d \alpha} (R_3(\alpha) \psi(\vec x))
= \frac{d}{d \alpha} \psi(R_3(\alpha)^{-1} \vec x)
= \frac{d}{d \alpha} \psi(x_1 \cos{\alpha} + x_2 \sin{\alpha},\ -x_1 \sin{\alpha} + x_2 \cos{\alpha},\ x_3) = \\
\displaystyle
= -(x_1 \cos{\alpha} + x_2 \sin{\alpha}) \frac{\partial \psi}{\partial x_2}
+ (x_2 \cos{\alpha} - x_1 \sin{\alpha}) \frac{\partial \psi}{\partial x_1}
\end{array}
$$
At $\alpha = 0$ we obtain
$$
\frac{d}{d \alpha} (R_3(\alpha) \psi(\vec x))|_{\alpha=0}
= - x_1 \frac{\partial \psi}{\partial x_2} + x_2 \frac{\partial \psi}{\partial x_1}
= -\left(\vec x \times \frac{\partial \psi}{\partial \vec x}\right)_3 = -l_3 \, \psi.
$$
So, in (\ref{rot}), substituting $R = R_i(\alpha)$, taking the derivative with respect to $\alpha$,
and substituting $\alpha=0$, we obtain the commutation relation (\ref{comm}):
$$
(l_i \circ F)\, \psi = -\frac{d}{d \alpha} \left((R_i(\alpha) \circ F)\, \psi\right)|_{\alpha=0}
=  -\frac{d}{d \alpha} \left((F \circ R_i(\alpha))\, \psi\right)|_{\alpha=0}
= (F \circ l_i)\, \psi.
$$

\subsection{Hamiltonian form of the momentum equations}

This observation enables us to change the places of operators in (\ref{momAv}):
\begin{equation}
\dot{\vec L} = -\frac{1}{2 \pi} \varepsilon \ \left(\vec l \circ F \, \psi \right)
 = -\frac{1}{2 \pi} \varepsilon \ \left(\vec L \times \frac{\partial} {\partial \vec L}\right) \, (F \, \psi)(\vec L).
\label{mom2}
\end{equation}
But it is easy to see that $\vec L \times \frac{\partial} {\partial \vec L}$ acts on a given function $G(\vec L)$
as taking its Poisson bracket with $\vec L$:
$$
\vec L \times \frac{\partial G} {\partial \vec L} = \{G, \vec L\},
$$
where the usual Poisson brackets for the components of the momentum are assumed:
$$
    \{L_i, L_j\} = \sum_k \varepsilon_{ijk} L_k.
$$
So, we obtain that the averaged equations for the momentum can be formulated in the following
{\it Hamiltonian form}:
\begin{equation}
\dot{\vec L} = \{\vec L, H(\vec L)\},
\label{mom3}
\end{equation}
where the Hamiltonian is the Funk transform of the deformation function:
$$
H(\vec L) = \frac{1}{2 \pi} F \left(\varepsilon \,  \psi \right).
$$


\section{Conclusion}

The reduced Hamiltonian system has $\vec L^2$ as its first integral (a Casimir function),
so the phase space is the 2D-sphere.
Being a Hamiltonian system with one degree of freedom, it is completely integrable.
The topological structure of the system's dynamics can be studied
by means of phase portraits on the sphere $\vec L^2 = const$.
The trajectories are the contour lines of the Hamiltonian.

It is important to note that, in the generic case when the Hamiltonian is a Morse function,
the topological characterization of the phase portrait can be performed using
the topological classification of Morse functions on the 2D-sphere developed by V.I. Arnold in \cite{Arn}.

The author is thankful to prof. V.L. Golo for the attention to this paper.

The work was supported by the grants RFFI 09-02-00551, 09-03-00779,
Grant for leading scientific schools NSH-3224.2010.1.


\begin{thebibliography}{99}

\bibitem{Jac} C.G. Jacobi, Vorlesungen \"uber Dynamik, Lec. 28, URSS, Moscow (2004).

\bibitem{ArnMech} V. I. Arnold, Mathematical Methods of Classical Mechanics, 2nd edition,
Springer-Verlag, New York (1995).

\bibitem{GS} V.L. Golo and D.O. Sinitsyn, Asymptotic Hamiltonian reduction for the dynamics
  of a particle on a surface, Physics of Particles and Nuclei Letters, Vol. 5, No. 3, pp. 278-281 (2008).

\bibitem{Gelf} I.M. Gelfand, S.G. Gindikin, M.I. Graev, Selected topics in integral geometry
(Translations of Mathematical Monographs),  American Mathematical Society (2003).

\bibitem{Arn} V.I. Arnold, Topological classification of Morse functions and generalisations of Hilbert's 16-th problem,
Math. Phys. Anal. Geom., Vol. 10, No. 3, pp. 227-236 (2007).



\end{thebibliography}
\end{document}